\def\ds{\displaystyle}
\def\ni{\noindent}
\def\ha{{1 \over 2}}
\def\fb{\overline f}
\def\gb{\overline g}
\def\Vb{\overline V}
\def\Xb{\overline X}
\def\Yb{\overline Y}
\def\Cb{\overline C}
\def\etab{\overline \eta}
\def\Ab{\overline A}
\def\Bb{\overline B}
\def\Kb{\overline K}
\def\di{\partial}
\def\rp{r^{\prime}}
\def\elb{\;{\overline{\! \ell}}}
\def\Fb{\overline F}
\def\Gb{\overline G}
\def\Ub{\overline U}

\magnification=\magstephalf
\baselineskip=18pt                      
\centerline{\bf INTEGRAL IDENTITIES  AND BOUNDS FOR SCATTERING CALCULATIONS }
\centerline{\bf IN THE DIRAC FORMALISM }
\vskip 1truecm
\centerline{Jurij W. Darewych }
\centerline{Department of Physics and Astronomy}
\centerline{York University, Toronto, ON  M3J 1P3 Canada}
\vskip 1truecm
{\bf Abstract}

{\narrower { Integral identities that hold between ``desired'' and ``comparison'' 
solutions of the radial Dirac equations for scattering precesses are 
considered. Applications of these identities are discussed, particularly the 
determination of bounds to variational calculations of $K$-matrix elements.}}

\vskip 1truecm

\par
Relativistic effects in atomic scattering, and indeed in many quantum scattering processes, 
can usually be treated as small corrections, and handled by perturbation theory with respect
 to the non-relativistic (Schr\"odinger) results.  However, for some processes, such as the 
scattering of electrons by atoms and molecules, and for many nuclear scattering processes,
 it is often convenient or even necessary to use the Dirac equation 
directly. This is because the relativistic kinematics and spin effects are then automatically ``built in'' 
(see, for example, ref. [1 - 3] and citations therein). 
 In addition to this, relativistic corrections to the 
dynamics ({\sl i.e.} corrections to the static potential) may need  to be taken into account. 

In the Dirac formalism, the description of the scattering of a fermion 
(such as an electron or positron)
 by a target (such as a neutral atom) is often reducible to the solution of the radial 
Dirac equations ($\hbar = c = 1$)
$$
f^{\prime}(r) + {\kappa \over r} f(r) = (E+m-V(r)) g(r) + X(r), \eqno (1)
$$
$$
g^{\prime}(r) - {\kappa \over r} g(r) = -(E-m-V(r)) f(r) + Y(r), \eqno (2)
$$
where $f(r)$ and $g(r)$ are the usual reduced radial coefficients of the ``large'' 
and ``small'' components of the Dirac spinor for the incident fermion of mass $m$ 
and energy $E$. 
The terms $X(r)$ and $Y(r)$ include exchange effects as may be applicable. 
We shall restrict our discussion to the class of potentials $V(r)$, and corresponding 
exchange terms $X(r), Y(r)$, such that the solutions $f(r), g(r)$ satisfy the following 
boundary conditions:
$$
f_\kappa(r=0) = g_\kappa(r=0) = 0, \eqno (3)
$$
$$
f_\kappa(r \to \infty) \sim A_\kappa(k) \sin \left(k r-\ell {\pi \over 2}\right) + 
B_\kappa(k) \cos \left(k r-\ell {\pi \over 2}\right), \eqno (4)
$$
$$
g_\kappa(r \to \infty) \sim  {f^{\prime}_\kappa(r) \over {E + m}}
 = {k \over {E+m}}\left( A_\kappa(k) \cos \left(k r-\ell {\pi \over 2}\right) - 
B_\kappa(k) \sin \left(k r-\ell {\pi \over 2}\right) \right), \eqno (5)
$$
where $k^2 = E^2 - m^2$, and $\kappa = j+\ha$ if $j=\ell-\ha$ (``spin down'')
 and  $\kappa = -(j+\ha)$ if $j=\ell+\ha$ (``spin up''). 
This means that the potentials are short range and not overly singular at the 
origin. Specifically, a sufficient condition is that the limit of $r^2 V(r)$ be 
zero as $r \to \infty$ and as $r \to 0$.
If the potential $V(r)$ is long-range, that is contains a Coulombic contribution, then 
the sine and cosine functions in (4) and (5) would be replaced by the corresponding 
Coulomb functions.

The asymptotic forms (4) and (5) can be written in the equivalent form
$$
f_\kappa(r \to \infty) \sim C_\kappa(k) \sin \left(k r-\ell {\pi \over 2}+ \eta_\kappa(k) \right)  
, \eqno (6)
$$
$$
g_\kappa(r \to \infty) \sim {{k } \over {E+m}} C_\kappa(k) \cos \left(k r-\ell 
 {\pi \over 2}+ \eta_\kappa(k) \right)  
, \eqno (7)
$$
where $ \eta_\kappa(k)$ are the scattering phase shifts, while 
$$ A_\kappa(k) =  C_\kappa(k) \cos  \eta_\kappa(k) \;\; {\rm and} \;\; 
 B_\kappa(k) =  C_\kappa(k) \sin  \eta_\kappa(k). \eqno (8)
$$

The asymptotic normalization constants $ C_\kappa(k)$ (or, equivalently, the constants 
  $A_\kappa(k), B_\kappa(k)$) may be chosen to be 
anything that is convenient. Some common choices are $ C_\kappa(k) = 1, 
 C_\kappa(k) = \sec  \eta_\kappa(k),$ etc.
The scattering cross sections or polarization parameters are then calculated 
from the phase shifts $ \eta_\kappa(k)$ [1,2].

	In non-relativistic (Schr\"odinger) scattering theory perturbative effects
can be taken into account by using the integral identity between a ``given'' 
and ``comparison'' solution first obtained by Hulth\'en [4] and later elaborated 
by Kato [5] and others. This integral identity can also 
serve as the basis for approximate variational solutions to the scattering equations [4-7], 
and for determining bounds on approximate calculations of scattering parameters [8,9].

Evidently, analogous results can be written down in the Dirac formalism of scattering 
theory, as we now proceed to discuss. Thus, 
suppose $\fb (r), \, \gb (r)$ are solutions of a ``trial'' or ``comparison'' problem, 
corresponding to  $\Xb (r)$, $\Yb (r)$ and the potential $\Vb (r)$, namely 
$$
\fb ^{\prime}(r) + {\kappa \over r} \fb (r) = (E+m-\Vb (r)) \gb (r) + \Xb (r), \eqno (9)
$$
$$
\gb ^{\prime}(r) - {\kappa \over r} \gb (r) = -(E-m-\Vb (r)) \fb (r) + \Yb (r), \eqno (10)
$$
with
$$
\fb_\kappa(r \to \infty) \sim \Cb_\kappa(k) \sin \left(k r-\ell {\pi \over 2}+ \etab_\kappa(k) \right)  
, \eqno (11)
$$
$$
\gb_\kappa(r \to \infty) \sim {{k} \over {E+m}}  \Cb_\kappa(k)  \cos \left(k r-\ell 
 {\pi \over 2}+ \etab_\kappa(k) \right)  
, \eqno (12)
$$

Straightforward manipulations of the equations (1), (2) and (9), (10) result in the identity
$$
{d \over {dr}} (\fb g - f \gb) = (V - \Vb) (f \fb + g \gb) + \Xb g - X \gb + \fb Y - f \Yb.
 \eqno (13)
$$
Integration of Eq. (13) leads to the result
$$
\left[ \fb (r) g(r) - f(r) \gb (r) \right]^R_0 = \int^R_0 dr \left[ 
(V - \Vb) (f \fb + g \gb) + \Xb g - X \gb + Y \fb - \Yb f  \right], \eqno (14)
$$
where $f = f(r)$, etc. in the integrand of Eq. (14). If we now make the replacements 
$f = \fb +(f-\fb)$, etc., Eq. (14) can be rewritten in the form
$$\eqalignno{
\Delta &=  \int^R_0 dr \left[ (V - \Vb) ( \fb^2 +  \gb^2) \right] + 
 \int^R_0 dr  (V - \Vb) \left[(f- \fb) \fb + (g-\gb) \gb\right ] \cr
 & + \int^R_0 dr \left[(\Xb-X) \gb  - (\Yb - Y) \fb\right]
 + \int^R_0 dr \left[\Xb (g - \gb) -  \Yb (f -\fb)\right], & (15) \cr }
$$
where
$$
\Delta = \left[ \fb (r) g(r) - f(r) \gb (r) \right]^R_0 \sim {k \over {E + m}} C \Cb 
\sin (\etab_\kappa - \eta_\kappa) =  {k \over {E + m}} (\Bb A - \Ab B),
 \eqno (16)
$$
and the symbol $\sim$ indicates that $R$ has been taken to be sufficiently large that 
the asymptotic forms (6), (7), (11) and (12) apply (we can take $R \to \infty$).

The integral identities (14) and (15) relate the phase shifts $\eta_\kappa$ (or, more generally, 
functions of these, such as the  $K$-matrix elements, $K_\kappa = \tan \eta_\kappa$, 
$T$-matrix elements, $T_{\kappa} = e^{i \eta_{\kappa}} \sin \eta_{\kappa}$, etc.) 
to the ``comparison'' phase shifts $\etab_\kappa$ (or corresponding functions thereof).  This is clear 
from the explicit form of $\Delta$ for given choice of asymptotic normalization, that is, choice of 
$A$ and  $B$ or alternatively $C$. For example if $C=\Cb=1$ then $\ds {\Delta = {k \over {E + m}}
\sin (\etab_\kappa - \eta_\kappa)}$, 
or if $A=\Ab=1, B=\tan \eta_\kappa=K_\kappa,   \Bb=\tan \etab_\kappa=\Kb_\kappa$ 
then $\ds {\Delta = {k \over {E + m}}(\Kb_\kappa - K_\kappa)}$, etc. 

The integral identities (14) or (15) can be used for various purposes, some of which we 
discuss briefly in what follows:

\vskip .5truecm
\ni 1. {\bf \sl Formal results.}

If we take $\Vb = 0$, and the corresponding free incident wave solutions of Eqs. (9) and (10) 
for $\fb_\kappa$ and $\gb_\kappa$, then the identity (14) (with  $A=\Ab=1, B=\tan \eta_\kappa
=K_\kappa, \Bb=0$)
 gives the well-known integral expression 
for the $K-$matrix elements,
$$
K_\kappa = - {{E+m}\over k} \int^R_0 dr \left[ 
V (f \fb + g \gb)  - X \gb + Y \fb   \right]. \eqno (17)
$$
 This is often used for extracting the phase shifts from numerical solutions of Eqs. (1) and (2).

\vskip .5truecm
\ni 2. {\bf \sl Perturbative calculations.}

A not-infrequent situation is that the potential $V$ can be written in the form $V=V_0 + V_1$, 
where $V_0$ 
is a dominant (and/or easily solvable) interaction term (such as the electrostatic potential in atomic 
scattering), and $V_1$ is a small ``correction'' term. Then, obviously, if $\Vb = V_0$, 
$\fb$ and $\gb$ are known (or easily obtainable), while $V-\Vb=V_1$ can be handled perturbatively.  
Taking $f=\fb$ and $g=\gb$ in lowest order on the right-hand-side of Eq. (15), 
one can use that equation to evaluate $\eta_\kappa$ in terms of $\etab_\kappa$ plus a lowest order 
perturbative correction (which is given by the RHS of Eq. (15) with $f=\fb$ and $g=\gb$).
 In general the perturbation may be in $V$ only, or in $X$ and $Y$, or both 
(see, for example, ref. [10]).

\vskip .5truecm
\ni 3. {\bf \sl Variational approximations.}

In some instances it may be useful or necessary to approximate the solutions of (1) and (2) 
variationally. For example, one may wish to have analytic representations of the solutions 
(recall that, with rare exceptions, Eqs. (1) and (2) are not analytically solvable). 
In such cases, one can use a 
variational approach, in which the desired (unknown) solutions $f(r),\, g(r)$ are approximated by 
analytic trial forms $\fb (r), \, \gb (r)$ that contain adjustable parameters $\alpha_j \;\; (j=1,...,n_p)$.
The identity (15) can be used to choose these parameters $\alpha_j$ in a variationally optimal 
way. We illustrate this on the case $X=Y=\Xb=\Yb=0$, and normalization choice $C_\kappa =
 \Cb_\kappa = 1$, in which case the identity (15) can be written as
$$
{k \over {E+m}} \sin (\etab_\kappa - \eta_\kappa) = I[\fb,\gb] + {\cal R}_2[f,g,\fb,\gb], \eqno (18)
$$
where
$$\eqalignno {
I[\fb,\gb]&=\int^R_0 dr (V-\Vb)(\fb^2+\gb^2) &(19a) \cr
&=\int^R_0 dr \left[ \fb \left ( ({d \over {dr}} - {\kappa \over r}) \gb + (E-m-V) \fb \right )
 -\gb \left (  ({d \over {dr}} + {\kappa \over r}) \fb - (E+m-V) \gb \right ) \right], & (19b) \cr
 } $$
and where we have used the identities
$$
 \left ( {d \over {dr}} + {\kappa \over r} \right) \fb (r) - \left(E+m-V(r)\right) \gb (r) 
=  \left( V(r)-\Vb (r) \right) \gb (r), \eqno (20)
$$
$$
\left({d \over {dr}} - {\kappa \over r}\right) \gb (r) + \left(E-m-V(r)\right) \fb (r) =
 - \left( V(r)-\Vb (r) \right) \fb (r). \eqno (21)
$$
in rewriting (19a) in the form (19b).
The term ${\cal R}_2$ is a ``remainder'' that is given by the expression
$$
{\cal R}_2[f,g,\fb,\gb] = \int^R_0 dr\,(V- \Vb) \left [(f-\fb)\fb + (g-\gb) \gb \right], \eqno (22)
$$
 which is second order in the ``small'' quantities $f-\fb, g-\gb,$ 
and  $V-\Vb$. 
 Usually we take $R \to \infty$ in these integrals, and this shall be done in the rest of this paper.

From Eq. (18), if we neglect ${\cal R}_2$, it follows that
$$
\eta_\kappa (k) \simeq \etab_\kappa (k) - {{E+m}\over k} \sin^{-1} I[\fb,\gb] = 
\eta_\kappa^{\rm (App.)} (k, \alpha_j), \eqno (23)
$$
where $\eta_\kappa^{\rm (App.)} (k, \alpha_j)$ is the approximate value of $\eta_\kappa (k)$ for any 
given $k$ and $\kappa$.  Note that an explicit knowledge of the comparison 
potential $\Vb$ is not necessary to evaluate $\eta_\kappa^{\rm (App.)} (k,\alpha_j)$, 
that is, it is only necessary to choose the trial functions $\fb,\, \gb$. This is evident from 
Eq. (19b), in which, as can be seen,  $\Vb$ does not appear explicitly.  
Of course, we want to choose the adjustable parameters $\alpha_j$ of $\fb(r,\alpha_j)$
 and $\gb (r, \alpha_j)$ in such a way that $\eta_\kappa^{\rm (App.)} (k, \alpha_j)$ is as 
close to $\eta_\kappa (k)$  as possible.  In other words, we wish to minimize 
$\left|\eta_\kappa (k) - \eta_\kappa^{\rm (App.)} (k, \alpha_j)\right|$ with respect to $\alpha_j$.
Since
$$
{\di \over {\di \alpha_j}} \left|\eta_\kappa (k) - \eta_\kappa^{\rm (App.)} (k, \alpha_j) \right| = 
- {\left( \eta_\kappa (k) - \eta_\kappa^{ \rm (App.)} (k, \alpha_j)\right) 
\over {\left|\eta_\kappa (k) - \eta_\kappa^{ \rm (App.)} (k, \alpha_j) \right|}} 
{\di \over {\di \alpha_j}}  \eta_\kappa^{\rm (App.)} (k, \alpha_j), \eqno (24)
$$
we see that a condition for a minimum of $ |\eta_\kappa (k) - \eta_\kappa^{\rm (App.)}
 (k, \alpha_j)|$ is that 
$$
{\di \over {\di \alpha_j}}  \eta_\kappa^{\rm (App.)} (k, \alpha_j) = 0, \eqno (25)
$$
as happens also in the corresponding Schr\"odinger theory.  The resulting optimal values,
 $\alpha_j^{\rm opt}$, of the adjustable parameters $\alpha_j$ (which include the trial value 
$\etab_\kappa$ of the phase shift, or of $\Kb_\kappa = \tan \etab_\kappa$ if the normalization
 $\Ab=1, \Bb=\tan \etab_\kappa$ is used, etc.), are then substituted 
into Eq. (23) to yield the optimal variational 
approximation,  $ \eta_\kappa^{\rm (App.)} (k, \alpha_j^{\rm opt})$, 
to $\eta_\kappa (k)$ (or $K_\kappa = \tan(\eta_\kappa)$, etc.), corresponding to 
a minimum value of $|{\cal R}_2|$. (Strictly speaking, minimuma of 
 $ |\eta_\kappa (k) - \eta_\kappa^{\rm (App.)}
 (k, \alpha_j)|$ may occur at points in parameter space where 
$\ds {{\di \over {\di \alpha_j}}  \eta_\kappa^{\rm (App.)} (k, \alpha_j) }$ is 
undefined (i.e. cusps rather than smooth minima) or at boundary points of the 
domain of parameter space.  Such possibilities must be kept in mind and investigated, 
if necessary.)

\vskip .5truecm
\ni 4. {\bf \sl Bounds on scattering parameters.}

In approximate calculations of scattering parameters (phase shifts, $K$-matrix elements, etc.) 
neither the sign nor the magnitude of the difference between the (unknown) exact and
approximate value is known. However, for the case $X=\Xb=Y=\Yb=0$, if we write 
$\Vb= V+\delta V,\, \fb= f+\delta f$ and $\gb = g + \delta g$, where $\delta V,\,  \delta f, 
\, \delta g \to 0$, then (with the choice of asymptotic normalization $C=1$) Eq. (14) 
implies that
$$
{k \over {E+m}} (\etab_\kappa - \eta_\kappa) = - \int_0^\infty dr\, \delta V (\fb^2 + \gb^2),
 \eqno (26)
$$
where we have kept only the first order terms in infinitesimal quantities, and so set 
$\sin (\delta \eta) = \delta \eta$. Equation (26) shows that if $\Vb \to V$ from below, i.e.
if $\delta V = \Vb - V < 0$, then $\etab_\kappa > \eta_\kappa$ and vice-versa 
(as happens also in Schr\"odinger theory). This property can be used to
set up a scheme in which  
approximate calculations of phase shifts approach the (unknown) exact values 
from above (or below), provided that the trial solutions are so chosen that the corresponding 
trial potentials approach the exact one from below (or above). 

Although, as already stated, in general it is not possible to evaluate either the sign or 
the magnitude of the 
remainder term ${\cal R}_2$, Eq. (22), it is possible, in some cases, to determine calculable 
bounds ${\cal B}$ on ${\cal R}_2$ of the form 
$$\left|{\cal R}_2[f,g,\fb,\gb]\right| < {\cal B}[V,\fb,\gb]. \eqno (27) $$
This, together with Eq. (18) (or its equivalent with other asymptotic normalizations),  
leads to upper and lower bounds on the scattering parameters. We illustrate this on the potential 
scattering case ($X=Y=\Xb=\Yb=0$), and the choice of  asymptotic 
normalization $A=\Ab=1, B=\tan \eta_\kappa= K_\kappa, \Bb=\tan \etab_\kappa 
= \Kb_\kappa$, whereupon Eq. (14) becomes
$$
{k \over {E+m}} (\Kb_\kappa - K_\kappa) = I[\fb,\gb] + {\cal R}_2, \eqno (28)
$$
where $ I[\fb,\gb]$ is given in Eq. (19).

We write the remainder term, Eq. (22) in the form
$$
{\cal R}_2 = {\cal R}_{2L} + {\cal R}_{2S}, \eqno(29)
$$
where
$$ 
{\cal R}_{2L} = \int_0^\infty dr \Delta V F(r) \fb (r),
\;\;\;\;\;\   {\cal R}_{2S} = \int_0^\infty dr \Delta V G(r) \gb (r),
\eqno(30) $$
with $\Delta V = V-\Vb$, $F=f-\fb$ and $G=g-\gb$.  Then, using the Schwartz 
inequality $(s,t)^2 \le (s,s) (t,t)$, it follows from Eq. (29) that
$$
|{\cal R}_{2L}| \le a_F b_{\fb}, \;\;\; |{\cal R}_{2S}| \le a_G b_{\gb} \;\;\;
{\rm and} \;\; {\rm  so} \;\;\; |{\cal R}_{2}| < a_F b_{\fb} +a_G b_{\gb}, \eqno (31)
$$
where
$$
b_{\fb}^2 = \int_0^\infty dr \rho^{-1}(r) |\Delta V(r) \fb(r)|^2, \;\;\; 
b_{\gb}^2 = \int_0^\infty dr \rho^{-1}(r) |\Delta V(r) \gb(r)|^2,  \eqno (32)
$$
$$
a_F^2 = \int_0^\infty dr \rho (r) |F (r)|^2,  \;\;\; 
a_G^2 = \int_0^\infty dr \rho (r) |G (r)|^2,    \eqno (33)
$$
and $\rho (r)$ is an arbitrary, positive weight function (but such that all the 
indicated integrals exist). For example, $\rho (r)$ might be chosen to be $|\Delta V(r)|$,
 or some other positive function, possibly with adjustable parameters, such that 
the indicated integrals exist.
For a given choice of trial functions $\fb,\; \gb$, the expressions $b_{\fb}$ and 
$b_{\gb}$ of Eq. (32) are calculable (remember that $\Vb$ need not be 
known explicitly, in light of the identities (20) and (21)).

	It remains now to determine bounds on $a_F$ and $a_G$ (which are 
not calculable since $F=f-\fb$ and $G=g-\gb$ are not known). One way that 
such bounds can be obtained is from the integral equations for the radial Dirac functions
(written here for the present choice of asymptotic normalization $A=\Ab=1$ 
so that $B=K=\tan \eta$ and $\Bb=\Kb=\tan \etab$):
$$
f_\kappa (r) = u_1 (r) +\int^\infty_0 d \rp U(\rp) \left[ G_\ell^{11} (r,\rp) f_\kappa (\rp) 
 + G_\ell^{12} (r,\rp) g_\kappa (\rp) \right], \eqno (34)
$$
$$
g_\kappa (r) = u_2 (r) +\int^\infty_0 d \rp U(\rp) \left[ G_\ell^{21} (r,\rp) f_\kappa (\rp) 
 + G_\ell^{22} (r,\rp) g_\kappa (\rp) \right], \eqno (35)
$$
where the Green functions $ G_\ell^{ab}$ are defined by
$$\eqalignno {
 G_\ell^{ab} (r,\rp)  &= {1 \over k} v_a (kr) u_b (k\rp) \;\;\; \rp < r \cr
	 &= {1 \over k} u_a (kr) v_b (k\rp) \;\;\; \rp > r & (36)  \cr        
} $$
and $U(r) = (E+m) V(r)$.  The functions $u_a,\, v_a$ are defined in terms of the usual 
Ricatti-Bessel and Ricatti-Neumann functions [11], 
$$
{\hat j}_\ell (kr) = kr j_\ell (kr) \sim \sin (kr- \ell {\pi \over 2}) \;\; {\rm  and} 
 \;\; {\hat n}_\ell (kr) = kr n_\ell (kr) \sim - \cos (kr- \ell {\pi \over 2}), \eqno (37)
$$
namely
$$
u_1 (kr) = {\hat j}_\ell (kr)  \;\;\;\; v_1(kr) = {\hat n}_\ell (kr), \eqno (38)
$$
$$
u_2 (kr) = \sigma _\kappa {k \over {E+m}} {\hat j}_{\elb} (kr)  
\;\;\;\; v_2(kr) =   \sigma _\kappa {k \over {E+m}} {\hat n}_{\elb} (kr), \eqno (39)
$$
where $\elb = \ell - \sigma_\kappa$ and $\ds {\sigma_\kappa = {\kappa \over {|\kappa|}}}$ 
is the sign of $\kappa$.
Similar integral equations can be written down for $\fb$ and $\gb$, hence also for 
$F = f-\fb$ and $G=g-\gb$, specifically
$$
F_\kappa (r) =\Fb_\kappa (r) +\int^\infty_0 d \rp U(\rp) \left[ G_\ell^{11} (r,\rp) F_\kappa (\rp) 
 + G_\ell^{12} (r,\rp) G_\kappa (\rp) \right], \eqno (40)
$$
$$
G_\kappa (r) = \Gb_\kappa (r) +\int^\infty_0 d \rp U(\rp) \left[ G_\ell^{21} (r,\rp) F_\kappa (\rp) 
 + G_\ell^{22} (r,\rp) G_\kappa (\rp) \right], \eqno (41)
$$
where
$$
\Fb_\kappa (r) =\int^\infty_0 d \rp \left(U(\rp)- \Ub(\rp) \right)
 \left[ G_\ell^{11} (r,\rp) \fb_\kappa (\rp) 
 + G_\ell^{12} (r,\rp) \gb_\kappa (\rp) \right], \eqno (42)
$$
$$
\Gb_\kappa (r) = \int^\infty_0 d \rp \left(U(\rp)- \Ub(\rp) \right)
 \left[ G_\ell^{21} (r,\rp) \fb_\kappa (\rp) 
 + G_\ell^{22} (r,\rp) \gb_\kappa (\rp) \right]. \eqno (43)
$$
Note that $\Fb_\kappa (r)$, $\Gb_\kappa (r)$ are known functions, for given trial 
functions $\fb_\kappa$ and $\gb_\kappa$, since $ G_\ell^{ab}(r,\rp)$ are known. 
We stress that the explicit form of the trial potential $\Ub (r)$ need not be known in 
Eqs. (42) and (43) because of the identities (20) and (21).
Thus, only the trial functions $\fb_\kappa (r,\alpha_j)$ and  $\gb_\kappa (r,\alpha_j)$ 
need be specified. 

Now, multiplying Eq. (42) by $\rho (r) F^*(r)$, integrating over $r$, and making repeated 
use of the Schwartz inequality gives the result
$$
a_F \le a_{\Fb} + a_F g_{11} + a_G g_{12}, \eqno (44)
$$
and similarly
$$
a_G \le a_{\Gb} + a_F g_{21} + a_G g_{22}, \eqno (45)
$$
where $a_{\Fb}$ and $a_{\Gb}$, defined as in Eq. (33), are calculable 
since $\Fb$ and $\Gb$ 
are known. The factors $g_{ij}$ are given by
$$
g_{ij}^2 = \int_0^\infty  \int_0^\infty dr d\rp \rho (r) |G_\ell^{ij} (r,\rp) 
U(\rp)|^2 \rho^{-1} (\rp) .  \eqno (46)
$$
The generalized Schwartz inequality
$$
\left| \int \int dr d\rp s(r) Q(r,\rp) t(\rp) \right|^2 \le \int dr |s(t)|^2 \int d\rp |t(\rp)|^2 
\int \int dr d\rp |Q(r,\rp)|^2, \eqno (47)
$$
was used in obtaining the results (44)-(46).

From Eqs. (44) and (45), it follows that
$$
a_F \le {1 \over {\cal D}} \left[ (1-g_{22}) a_{\Fb} + g_{12} a_{\Gb} \right] = B_F,
 \eqno (48)
$$
$$
a_g \le {1 \over {\cal D}} \left[ (1-g_{11}) a_{\Gb} + g_{21} a_{\Fb} \right] = B_G,
 \eqno (49)
$$
provided that
$$
g_{ii}< 1 \;\; (i=1,2) \;\;\;\; {\rm and}\;\;\;\;
 {\cal D} =  (1-g_{11})  (1-g_{22}) - g_{12}  g_{21} > 0. \eqno (50)
$$
From the definition (46) of $g_{ij}$, and that of the Green functions (36), it is clear that 
the conditions (50) are, for given k (i.e. given energy of incidence), restrictions on the 
strength of the potential $V(r)$. That is, the potential must be sufficiently weak for the 
inequalities (50) to be met.  Note, however, that since the Green functions contain the 
factor $\ds {1 \over k}$, $g_{ij}$ will generally decrease with increasing $k$.  This means that 
a given potential $V(r)$ may be such that the inequalities (50) might not hold when $k$ is 
small (low-energy scattering) but will hold for higher values of $k$.

Replacing the expressions $a_F$ and $a_G$ by their bounds (48) and (49) in Eq. (31) then  
leads to the inequality
$$
|{\cal R}_2| < B_F b_{\fb} + B_G  b_{\gb} = {\cal B}_2,  \eqno (51)
$$
and hence to the following simultaneous upper and lower bounds on the (unknown) 
exact $K$-matrix element:
$$
K_{\kappa}^{\rm (App.)} - {{E+m} \over k} {\cal B}_2 < 
K_{\kappa}^{\rm (Exact)} = \tan \eta_{\kappa} < 
K_{\kappa}^{\rm (App.)} + {{E+m} \over k} {\cal B}_2, \eqno (52)
$$
where
$$
K_{\kappa}^{\rm (App.)} = \Kb - {{E+m} \over k} I[\fb,\gb].  \eqno (53)
$$
The definition (53) is the analogue of that of Eq. (23) for the present choice 
of asymptotic normalization $A=\Ab=1$. 
Note, again, that the bounds (52) hold provided that all the integrals that 
enter into the expressions for $K_{\kappa}^{\rm (App.)}$ and   ${\cal B}_2$
 exist, and that the inequalities (50) apply.

We stress that the bound ${\cal B}_2$ of Eq. (51) (with Eqs. (32), (48) and (49)) 
 is expressible in terms of $V(r)$ and the trial functions 
$\fb_\kappa (r,\alpha_j)$ 
and  $\gb_\kappa (r,\alpha_j)$, hence it is ultimately a function of the adjustable parameters, 
that is ${\cal B}_2 (\alpha_j)$. These parameters may be chosen in accordance with the 
variational prescription (25), or such that the upper and lower bounds are as close as 
possible, i.e. such that ${\cal B}_2 (\alpha_j)$ is a minimum.  These two prescriptions are not 
the same but, for sufficiently flexible trial functions, they will yield similar results. In practice, 
the prescription (25) is simpler to implement. In either case ${\cal B}_2$ can be made as 
small as desirable (in the domain where the conditions (50) hold), provided that $\fb$ and 
$\gb$ are sufficiently flexible.

To summarize, we have presented integral identities that hold between  given and a comparison 
(or ``trial'') solutions for scattering calculations in the Dirac formalism.  
Various applications of these integral identities have been discussed, including their use in
approximate, variational solutions of the scattering parameters (phase shifts or functions thereof).
In particular, we have used these integral identities to establish rigorous and calculable 
bounds on the difference between the exact and approximate $K$-matrix elements 
for a wide class of potentials.
These bounds can be made as tight as necessary, if sufficiently flexible trial functions are 
used.  
 
	The financial support of the Natural Sciences and Engineering Research Council 
of Canada for this work is gratefully acknowledged.


\vfill \eject

{\ni \bf References}
\medskip
\par

\item{1.}J. Kessler, {\sl Polarized Electrons}, 2nd ed., Springer V., 1985.    
\par\noindent
\item{2.} M. E. Rose, {\sl Relativistic Electron Theory}, Wiley, 1961.  
\par\noindent
\item{3.} R. P. McEachran and A. D. Stauffer, {\sl Proceedings of the International  
Symposium on Correlation and Polarization in Electronic and Atomic Collisions}, A. Crowe 
and M. R. H. Rudge, eds., World Scientific, Singapore, 1988, p. 183.
\item{4.}L. Hulth\'en, Kgl. Fysisgraf. S\"altshap. Lund. Forn. {\bf 14}, 2 (1944). 
\par\noindent
\item{5.} T. Kato, Prog. Theor. Phys. {\bf 6}, 245 (1951), {\bf 6}, 394 (1951).
\par\noindent
\item{6.}  W. Kohn, Phys. Rev.  {\bf 74}, 1763 (1948).
\par\noindent
\item{7.} R. G. Newton, {\sl Scattering Theory of Waves and Particles}, 2nd ed., Springer V., 
New York, 1982.  
\par\noindent
\item{8.}J. W. Darewych and I. Schlifer, Can. J. Phys. {\bf 68}, 1179 (1990).
\par\noindent
\item{9.}J. W. Darewych and I. Schlifer,  J. Math. Phys. {\bf 33}, 2557 (1992).
\par\noindent
\item{10.} W. R. Johnson and C. Guet,  Phys. Rev. A {\bf 49}, 1041 (1994).
\par\noindent
\item{11.} H. M. Abramowitz and I. A. Stegun, {\sl Handbook of Mathematical 
Functions}, Dover, New York, 1965, Ch. 10.

\bye